# Database and data structure for the novel TOF-PET detector developed for J-PET project

E. Czerwiński[1], M. Zieliński[1], T. Bednarski[1], P. Białas[1], Ł. Kapłon[1,2], A. Kochanowski[2],
G. Korcyl[1], J. Kowal[1], P. Kowalski[3], T. Kozik[1], W. Krzemień[1], E. Kubicz[1], M. Molenda[2], P. Moskal[1],
Sz. Niedźwiecki[1], M. Pałka[1], M. Pawlik[1], L. Raczyński[3], Z. Rudy[1], P. Salabura[1], N.G. Sharma[1],
M. Silarski[1], A. Słomski[1], J. Smyrski[1], A. Strzelecki[1], A. Wieczorek[2], W.Wiślicki[3], N. Zoń[1]

[1]Faculty of Physics, Astronomy and Applied Computer Science,
Jagiellonian University, 30-059 Cracow, Poland
[2]Faculty of Chemistry, Jagiellonian University, 30-060 Cracow, Poland
[3]Świerk Computing Centre, National Centre for Nuclear Research, Otwock-Świerk, Poland

**Abstract:**
The complexity of the hardware and the amount of data collected during the PET imaging process require application of modern methods of efficient data organization and processing. In this article we will discuss the data structures and the flow of collected data from the novel TOF-PET medical scanner which is being developed at the Jagiellonian University. The developed data format reflects: registration process of the gamma quanta emitted from positron-electron annihilation, Front-End Electronic (FEE) structure and required input information for the image reconstruction. In addition, the system database fulfills possible demands of the evolving J-PET project.

**Introduction:**
Presently used and novelly developed Positron Emission Tomography (PET) scanners [1-8] are complex devices built from hundreds of small scintillating detectors which register big amount of data which needs to be handled and processed. For the J-PET project a new framework was developed [9] in order to control data processing and reconstruction. Such framework can run in the environment which provides sufficient data capacity, read-write speed and CPU power [10]. In order to optimize speed for both reconstruction code development, and data analysis a new data structure and database were created.

**Definitions – from $e^+e^-$ annihilation to input for image reconstruction:**
Single event of one positron-electron annihilation can be reconstructed in J-PET system only when 2 hits from created gamma quanta were registered exactly in two scintillator strips. Light emitted in the scintillator from one of these hits will travel along the strip and can be registered by photomultipliers (PM) attached to its ends. Analog signal from PM is probed at defined number of thresholds ($N_{th}$) by Front-End Electronics (FEE) [11,12] together with integrated charge of the signal, and stored in binary format. In the ideal case one $e^+e^-$ annihilation gives 2 hits in scintillator strips which then are registered as 4 signals in photomultipliers. Finally, single annihilation event is stored at $4\times(N_{th}+1)$ channels. The FEE developed for the J-PET project allows for trigger-less mode of data processing [11]. Time of all registered signals is measured with respect to the clock distributed among all electronic elements, and therefore time registered at specific channel (when signal crosses given threshold) is related to a time slot.

**J-PET basic elements:**
Scintillator strips are used to register gamma quanta in the described tomograph device. Light collected at the end of each strip is converted to electric signal by attached photomultiplier. Signals from several PMs attached to single FEE board are distributed to $N_{th}+1$ channels. Map of connections between scintillator-PM-FEE elements is stored in the dedicated database which allows to relate single output channel from FEE to a given threshold or charge of a signal registered by specific PM connected to known scintillator strip. Schematic view of the connection between J-PET basic elements is presented in Fig. 1.

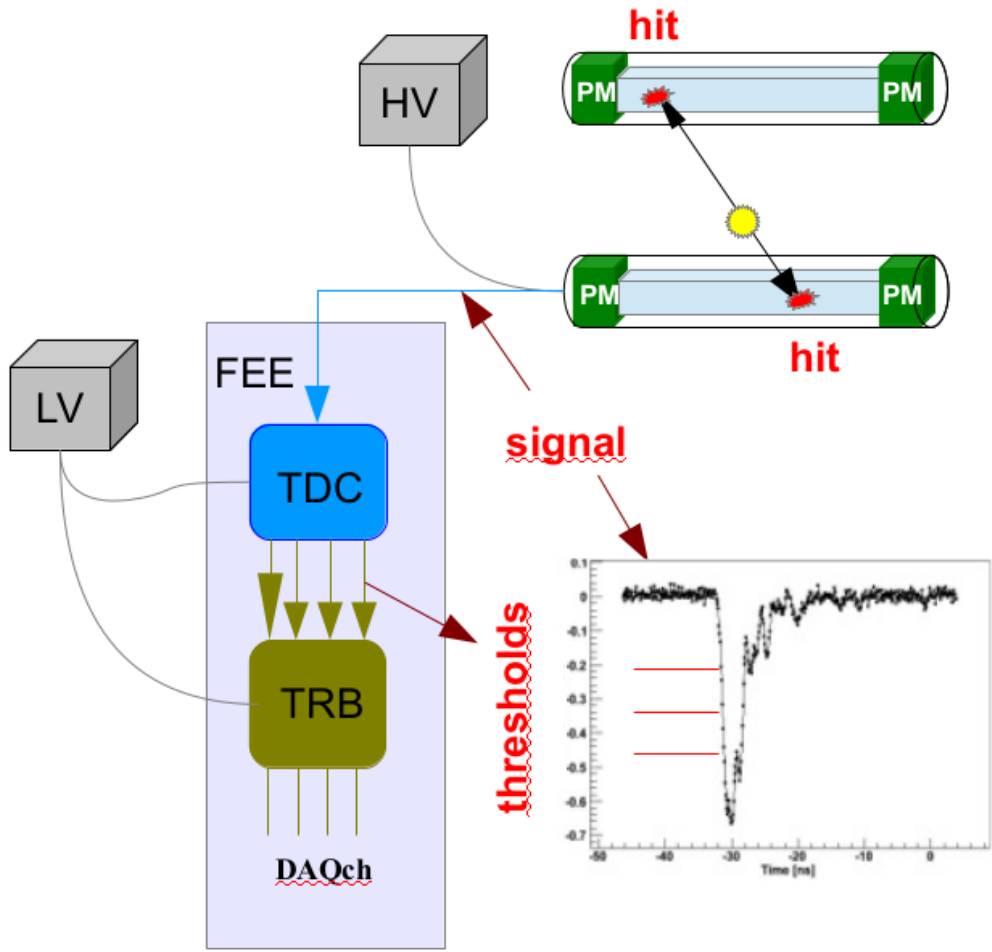

**Fig. 1** Schematic view of the annihilation event (yellow dot) and two hits of gamma quanta registered in two scintillating strips at J-PET. Light from each hit is converted to electronic signal by two photomultipliers. At given example one signal is probed by FEE at three thresholds and additionally information about charge of the signal is registered. For simplicity only two scintillator strips are shown and electronics for only one PM is presented.

**Reconstruction steps as data structure:**

Analysis steps of the J-PET data inside a developed framework corresponds to detector-FEE structure and physical process of annihilation event registration. Firstly binary data from FEE are unpacked to RootTree format [13]. In the next step the data are ordered into the structure which reflects detection system and FEE connections based on the information about current J-PET setup from the database. At this stage data can be corrected with calibration constants. Example of time calibration procedure is described in reference [14], while photomultiplier gain determination can be found in [15]. After applying time calibration constants a processed data can be converted from time slot to the true time of the event, which allows to group registered values at different channels into separate signals. Determination of time and position of the hit is based on the paired signals and reconstruction method [16,17]. Finally paired hits constitute input for further image reconstruction [18-20].

**J-PET Database design and structure**

For the medical records of the patient treatment all information about measurement conditions should be preserved. One needs to store all the information about the hardware used in the measurement, alignment of the detectors, initial setup parameters, calibrations, and the software configuration. In addition all these settings have to be saved for each single measurement, and available at any time for off-line data processing and analysis as well as for the backward compatibility.

To comprehend the majority of the different kind of data we developed and implemented the object-relational database model based on the open-source PostgreSQL engine [21] which is an object-relational database management system (ORDBMS). It features a very high stability, flexibility, and scalability. Furthermore, it provides variety of extensions for different systems, data types, operators, methods, views, aggregates, and procedural languages.

Logically developed database is organized in three main parts:
1. information about hardware properties and operational parameters like maximal high voltage values for the photomultipliers,
2. actual configuration of the setup, detector alignment and parameters like thresholds set on the front-end-electronics.
3. settings for each single measurement, as a readable log for the purpose of later off-line analysis of the data.

The logic structure of the database is shown in Fig. 2.

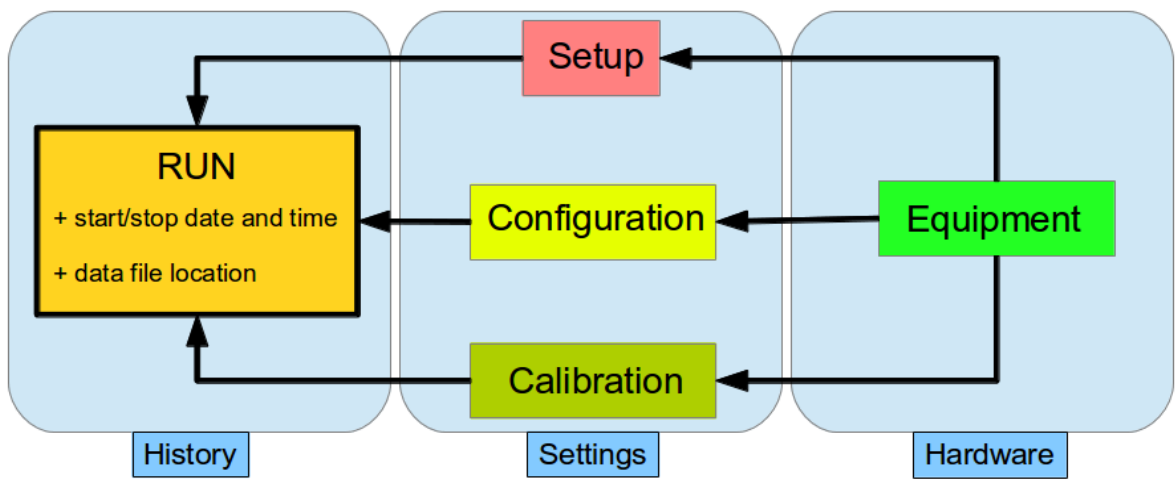

**Fig. 2** Schematic view of the implemented database logic. At given schema three logic parts of the database are seen: Hardware, Settings, and History.

Information about the hardware in the database contains inventory list about all the equipment and its major properties, which is used and available for the project of the novel PET device. It holds the details of: high and low voltage suppliers (numbers of output channels), scintillator strips (dimensions), photomultipliers (maximal value of high voltage for safety purposes), FEE (number of input, output channels), types of radiation sources and phantoms. Each single part of equipment has its own unique identification number which allows for easy identification of the hardware. For each hardware part the *status* value can be set to distinguish the situation for the *broken*, *terminated*, *available* and *in use* element. This ensures that the part which is e.g. broken will not be used in current setup by accident. Furthermore all the operations like adding, deleting, and changing of the status of particular equipment are logged per user and also stored in the database.

Moreover, some of the hardware: photomultipliers, scintillators and Front-End-Electronics, needs an initial calibration before usage in the measurement. In the developed database a separate unit within the hardware part is responsible to hold the calibration constants. This solution ensures the possibility for the change of the calibration at any time in the future and also stores the history of previously used values.

For the real measurement with phantoms and in the future for patient examination the hardware has to be assembled together and connected physically and logically. This structure of connections is mapped into the software as a *setup*. Every connection in the setup is represented by a single entity in the database and it references to two setup-objects. Firstly the connections between the HV channels, photomultipliers and scintillators are established. Later a sequence of FEE channels is defined, therefore each setup contains an information abut connection configuration. Configuration is created for single connection between two sides of a setup-objects and holds the specific information for the connection like e.g. threshold set on specific channel in the FEE board. In the process of setting up the connection and configuration between particular elements, a simple logic mechanisms are implemented which ensures i.e. impossible to set a high voltage above safety limit stored for that element. Beside the connections every setup references also the time/date properties and stores the information about the user who created the setup. It features as well the possibility of cloning the setup which allows for very easy modifications of present setup and keeping the old one in the history.

Apart from the setup, each conducted measurement has to be stored with the information containing time and date, description, path to collected data, used setup and the optimal calibration. This functionality is implemented as a *Run* table in the database, and containing all the important details of each single measurement. Utilizing this feature one can perform off-line data analysis and, if needed, restore every specific parameter used during the measurement.

For the database handling and user usage, a native application written in object oriented C# language [22] was developed. It offers all the features of modern graphical user interface (GUI) assuring the easy access to all the units of the database, enabling adding, deleting, updating and modification option of each hardware, setup/connection or run sections.

**Data flow and data storage**

In the presently developed PET prototype annihilation events are registered by the scintillators and photomultipliers attached to its ends, from which signals are gathered by FEE boards and, then DAQ system saves them to disk. Data collected in this way are then used for the off-line analysis and medical image reconstruction. Therefore, admittance to collected data should be fast and efficient with possibility of secure access from external locations. For that purpose we have developed a way of secure data storage and handling using local resources and external data storage. The work-flow of the data after it is being processed by the DAQ is presented in Fig. 3.

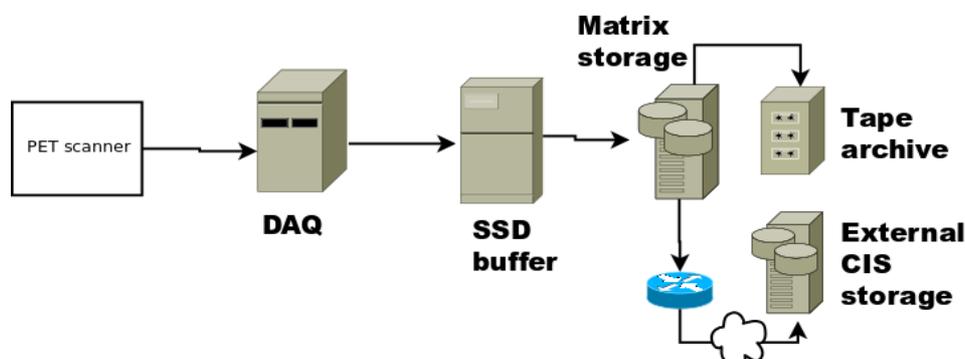

**Fig. 3** Schematic view of the implemented data flow and storage. Data collected from the PET scanner by DAQ system will be saved to data matrix and preserved on external CIS storage. Additional backup will be done with the tape archive.

The amount of collected data will be approximately few MB/s, which will be then saved in the files of size not exceeding 10GB. Assuming the measurements performed 40h/week and 52weeks/year, this makes around 100TB data collected per year. Data coming directly from the DAQ system will be saved to a fast SSD buffer storage with 2TB capacity. Later on it will be moved to temporary storage from which once per day data will be sent to a permanent cloud data storage in Świerk Computing Centre (CIS) [23]. For the purpose of the off-line analysis one will need to request specific data sample from the CIS cloud storage which will be than securely downloaded to a data matrix, coupled with a computing cluster used for the analysis. Additionally, for security reasons it is planed to preserve data collected during every measurement using the tape archive.

In the future the work-flow of the data will be organized based on the cloud and grid computing as presented in [10] to enable the optimal data saving, processing and fast medical image reconstruction.

**Summary:**

Presented structure of data and database are now working solutions in a last phase of development. The database structure is already tested and running. In addition the database graphical interface is being tested now. The system is designed to keep all settings during the measurement at development stage and in the future with small modifications as a working solution for J-PET detector.

**Acknowledgements**

We acknowledge technical and administrative support by M. Adamczyk, T. Gucwa-Ryś, A. Heczko, M. Kajetanowicz, G. Konopka-Cupiał, J. Majewski, W. Migdał, A. Misiak and the financial support by the Polish National Center for Development and Research through grant INNOTECH-K1/IN1/64/159174/NCBR/12, the Foundation for Polish Science through MPD programme and the EU and MSHE Grant No. POIG.02.03.00-161 00-013/09.

**References:**

[1] J.L. Humm, A. Rosenfeld, A. Del Guerra: *From PET detectors to PET scanners*, Eur. J. Nucl. Med. Mol. Imaging 30 1574 (2003)

[2] M. Conti: *State of the art and challenges of time-of-flight PET*, Phys. Med. 25 1-11 (2009)

[3] D.W. Townsend: *Multimodality imaging of structure and function*, Phys. Med. Biol. 53 R1–R39 (2008)

[4] J.S. Karp, et al.: *Benefit of Time-of-Flight in PET: Experimental and Clinical Results*, J. Nucl. Med. 49 462 (2008)

[5] D.W. Townsend: *Physical Principles and Technology of Clinical PET Imaging*, Ann. Acad. Med. Singapore 22 133 (2004)

[6] P. Moskal, P. Salabura, M. Silarski, J. Smyrski, J. Zdebik, M. Zieliński: *Novel detector systems for the Positron Emission Tomography*, Bio-Algorithms and Med-Systems 14, Vol 7, No. 2, 73, (2011); arXiv: 1305.5187

[7] P. Moskal, et al.: *Strip-PET: a novel detector concept for the TOF-PET scanner*, Nuclear

Medicine Review 15 (2012) C68; arXiv: 1305.5562

[8] P. Moskal, et al.: *TOF-PET detector concept based on organic scintillators*, Nuclear Medicine Review 15 (2012) C81; arXiv: 1305.5559

[9] W. Krzemień, et al.: *J-PET analysis framework for the prototype TOF-PET detector*, Bio-Algorithms and Med-Systems (2014) in print, arXiv: 1311.6153

[10] W. Wiślicki, et al.: *Computing support for advanced medical data analysis and imaging*, Bio-Algorithms and Med-Systems (2014) in print , arXiv:1401:6929

[11] G. Korcyl, et al.: *Trigger-less and reconfigurable data acquisition system for positron emission tomography*, Bio-Algorithms and Med-Systems (2014) in print

[12] M. Pałka, et al.: *A novel method based solely on FPGA units enabling measurement of time and charge of analog signals in Positron Emission Tomography*, Bio-Algorithms and Med-Systems (2014) in print , arXiv:1311:6127

[13] A data analysis framework, http://*root.cern.ch*

[14] M. Silarski, E. Czerwiński, et al., *A novel method for calibration and monitoring of time synchronization of TOF-PET scanners by means of cosmic rays*, Bio-Algorithms and Med-Systems (2014) in print, arXiv: 1311.6152

[15] T. Bednarski, E. Czerwiński, et al.: *Calibration of photomultipliers gain used in the J-PET detector*, Bio-Algorithms and Med-Systems (2014) in print, arXiv: 1312.2744

[16] L. Raczyński, et al.: *Novel method for hit-position reconstruction using voltage signals in plastic scintillators and its application to the Positron Emission Tomography*, Bio-Algorithms and Med-Systems (2014) in print

[17] L. Raczyński, et al.: *Application of Compressive Sensing Theory for the Reconstruction of Signals in Plastic Scintillators*, Acta Phys. Polon. B Suppl. 6 1021-1027 (2013), arXiv: 1310.1612

[18] A. Słomski, et al.: *3D PET image reconstruction based on MLEM algorithm*, Bio-Algorithms and Med-Systems (2014) in print

[19] P. Białas, et al.: *List mode reconstruction in 2D strip PET*, Bio-Algorithms and Med-Systems (2014) in print

[20] P. Białas, et al.: *System Response Kernel Calculation for List-mode Reconstruction in Strip PET Detector*, Acta Phys. Polon. B Suppl. 6 1027-1036 (2013), arXiv: 1310.1614

[21] website: http://www.postgresql.org

[22] ECMA C# Language Specification, ECMA- 334, http://www.ecma-international.org/publications/files/ECMA-ST/Ecma-334.pdf

[23] website: http://www.cis.gov.pl